\documentclass[letterpaper,prb,twocolumn]{revtex4}

\usepackage{amsmath}
\usepackage{fancyhdr}
\usepackage{graphicx}
\usepackage{subfigure}

\begin{document}
\textit{American Journal of Physics} \textbf{78}, 56-63 (2010)

\title{Modeling excitable systems: Reentrant tachycardia}

\author{Jarrett L. Lancaster}
\altaffiliation[Currently at ]{New York University, Physics Department, 4 Washington Place, New York, NY 10003}
\affiliation{University of North Carolina Greensboro, Department of Physics and Astronomy, Greensboro, North Carolina 27402}

\author{Esther M. Leise}
\affiliation{University of North Carolina Greensboro, Department of Biology, Greensboro, North Carolina 27402}

\author{Edward H. Hellen}
\email{ehhellen@uncg.edu} 
\affiliation{University of North Carolina Greensboro, Department of Physics and Astronomy, Greensboro, North Carolina 27402}


\begin{abstract}
Excitable membranes are an important type of nonlinear dynamical system and their study can be used to provide a connection between physical and biological circuits. We discuss two models of excitable membranes important in cardiac and neural tissues. One model is based on the Fitzhugh-Nagumo equations and the other is based on a three-transistor excitable circuit. We construct a circuit that simulates reentrant tachycardia and its treatment by surgical ablation. This project is appropriate for advanced undergraduates as a laboratory capstone project, or as a senior thesis or honors project, and can also be a collaborative project, with one student responsible for the computational predictions and another for the circuit construction and measurements.
\end{abstract}

\maketitle
\chead{}
\lhead{Accepted for publication in \textit{American Journal of Physics}}
\rhead{\thepage}
\section{Introduction}
Many physics majors are unaware of the important role that physicists play in medical research. Typical coursework for a physics major is removed from real-world applications, especially compared to what they perceive for their pre-medical peers. As a result, some students question whether majoring in physics will allow them to pursue a useful and satisfying career. These students will benefit from participation in projects that make connections between physical concepts and real-world medical problems. In this paper we focus on the modeling of reentrant tachycardias as a means to introduce physics majors to the physical underpinnings of excitable tissues.

The voltage changes that occur across cell membranes during action potentials (the signal pulses) in cardiac and neural tissues constitute an important type of excitable nonlinear behavior. Excitable systems typically display a threshold-triggered fast-response positive feedback followed by a slow-response inhibition that causes the system to return to its resting values. The recovery includes a phase known as the refractory period during which additional excitation is either more difficult or impossible to elicit. Excitation is manifested by the temporal pulse shape of a system variable such as the membrane potential during action potentials in cardiac and neuronal cells.\cite{Silverthorn} Another example at the cellular level is the calcium waves that occur inside cells during critical developmental events such as fertilization or changes in motility. Slower examples include changes in population dynamics that occur during disease epidemics and the ignition of vegetation in forest fires. Students can use theoretical and electronic excitable systems to explore the dynamics of these situations and test proposed control mechanisms such as cardiac defibrillation, vaccine intervention, and fire-fighting strategies. In this paper we model reentrant tachycardia responsible for atrial flutter, a common cardiac arrhythmia, and its treatment with surgical ablation (destruction of the abnormal tissue). 

One approach is to start with a mathematically simple excitable system and then build its electronic circuit realization. For this approach we use the Fitzhugh-Nagumo equations.\cite{Fitzhugh,Nagumo} Another approach is to start with a simple circuit that exhibits excitable behavior; we use a three transistor excitable circuit. We then find its mathematical representation. In both cases the models are presented in the form of an excitable system with excitatory variable $u$ and inhibitory variable $v$;
\begin{subequations}
\label{general}
\begin{align}
\frac{du}{dt}&=f(u,v)\\
\frac{dv}{dt}&=\epsilon \,g(u,v).
\end{align}
\end{subequations} 
Excitation pulses are supported by the reverse $N$-shape of the $u$-nullcline, $f(u,v)=0$ in phase space, and by $\epsilon \ll1$ causing a slow response of $v$.

Bunton et al.\cite{Bunton} describe an excitable circuit based on comparators and multiplexer integrated circuits and review other circuit models. Part of the appeal of their circuit is its simplicity, whose behavior can be understood by the action of a four-position switch without the use of equations. In this paper we emphasize using models in the form of Eq.~\eqref{general} to predict the behavior of circuits. The level of mathematics is similar to that found in current investigations into cardiac arrhythmias.\cite{Yuan,Cassia,Starobin} Numerical solution of the differential equations is easy with today's desktop computers using Python or many other popular programming environments. The projects described here are intended to introduce the behavior of excitable membranes and their actions in cardiac and neural systems. The projects also provide valuable applications of computational physics and circuit analysis and construction. 

Under normal conditions, action potentials that regulate heartbeat travel through the cardiac conduction system in a unidirectional fashion from the sinoatrial node to the atrioventricular node, through the atrioventricular bundle of fibers and eventually through the purkinje fibers in the wall of the ventricles.\cite{Silverthorn,Ross} This cardiac conduction system can malfunction and lead to a condition known as reentrant tachycardia, in which abnormally rapid heartbeats circulate through the conduction system without stopping. Several types of medically important tachycardias exist. The behavioral effects include shortness of breath, dizziness, fainting, heart palpitations, and chest pain. Tachycardias may occur in the absence of any obvious disease conditions, or may result from an overactive thyroid, electrolyte imbalance, or prior myocardial infarction (heart attack). Tachycardias may be treated with drugs, physical interventions, or in serious conditions, ablations of the abnormal tissue.\cite{Patel} 

In Sec.~II we briefly describe the fundamentals of excitations in cardiac and neural tissues and then discuss the Fitzhugh-Nagumo equations and associated circuit. In Sec.~III we present the three-transistor excitable circuit followed by its two-variable model. Section~IV gives an example of modeling a common cardiac arrhythmia, reentrant tachycardia, and its treatment by surgical ablation.

\section{Biological Excitability and the Fitzhugh-Nagumo Model}
In neural and cardiac tissue the momentary excitation pulses of the membrane potential (voltage difference from interior to exterior across the cell membrane), known as action potentials, are caused by ionic currents that flow through channels in the cell membrane. Fast-response positive feedback arises from the increased conductance through sodium and/or calcium ion channels that allow those ions to flow into the cell, thereby depolarizing or reducing the membrane potential. (The resting membrane potential is negative, so an inward current depolarizes the membrane by raising the potential toward zero.) In cardiac cells calcium ion currents cause the cardiac action potentials to have longer durations than most neuronal action potentials. The slow inhibitory response is caused by a delayed opening of potassium channels that allow those ions to flow out of the cell, thereby causing the recovery or repolarization (hyperpolarization) phase. A slowly responding decrease in the conductance of the sodium channels also adds to the recovery phase. The numbers and types of ion channels vary depending on the particular type of cell. The types of ion channels are distinguished by their voltage-gated threshold value, kinetics, and ion selectivities. Various ion pumps act like batteries to maintain ion concentration gradients across the cell membrane. 

The Hodgkin and Huxley model for the neuronal membrane potential accounts for many of the details of the sodium and potassium channel activities,\cite{HH} and accurately reproduces the shape of the action potential, at the expense of involving four variables. The two-variable Fitzhugh-Nagumo model is simpler and has excitable dynamics similar to the Hodgkin and Huxley model. As a result, Fitzhugh-Nagumo and similar models are popular in studies of cardiac and neuronal dynamics. 

\begin{figure}[h]
	\centering
		\includegraphics[angle=0,width=0.45\textwidth]{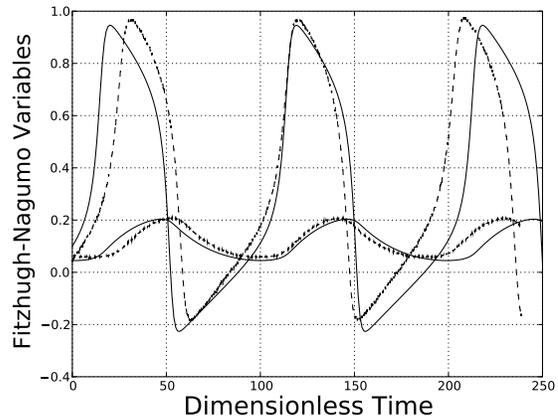}
	\caption{\label{fig:fn_time} Numerical computation (lines) and measured circuit data (dashed) for a self-exciting Fitzhugh-Nagumo system with source term $s(t)=0.06$. The excitatory variable $u$ has a large amplitude, and the inhibitory variable $v$ has a small amplitude. The voltages have been normalized and the time $t$ is dimensionless (see Table~\ref{tab:conv}).}
\end{figure}

The Fitzhugh-Nagumo equations expressed in the form of Eq.~\eqref{general} are\cite{Fitzhugh,Nagumo}
\begin{subequations}
\label{fn_eqns}
\begin{align}
\label{fn_eqns_a}
\frac{du}{dt}&=u(u-a)(1-u)-v+s(t)\\
\label{fn_eqns_b}
\frac{dv}{dt}&=\epsilon(u-bv).
\end{align}
\end{subequations}
Typical values of parameters that result in excitable behavior are $a=0.15$, $\epsilon = 0.01$, and $b=2.5$. The threshold-triggered positive feedback is due to $du/dt$ becoming positive for $a<u<1$ while $v$ and $s(t)$ are both zero. The slow response of the inhibitory variable $v$ is ensured by the small value of $\epsilon$. The function $s(t)$ is a source term. For example, setting $s(t)=0.06$ results in self-stimulating periodic excitation pulses as shown in Fig.~1. Alternatively, a time dependent source $s(t)$ such as a pulse train can be used to stimulate excitation pulses. 

\begin{figure}[h]
	\centering
		\includegraphics[angle=0,width=0.45\textwidth]{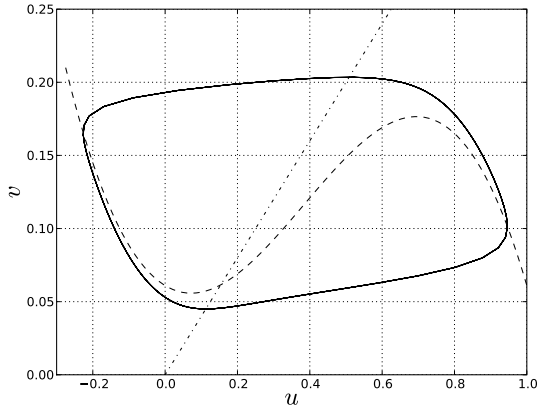}
	\caption{\label{fig:fn_phase}Numerical computation of phase-space plot (line) and nullclines ($u$-dashed, $v$-dot-dash) for the self-exciting Fitzhugh-Nagumo system.}
\end{figure}

Figure 2 shows the phase space trajectory of the excitation pulses in Fig.~1 and the nullclines of Eq.~\eqref{fn_eqns}. The $u$-nullcline is nonlinear and has the reverse-$N$ shape characteristic of systems that support excitation pulses. Fixed points occur at the intersection of the nullclines where the time derivatives of $u$ and $v$ are both zero. For $s(t)=0.06$ the fixed point is unstable resulting in the cyclic phase space trajectory. For $s(t)=0$ the nullclines intersect at the origin, and thus it is easy to use the Jacobian of Eq.~\eqref{fn_eqns} in a linearization stability analysis to show that the origin is a stable fixed point characterized as a spiral node.\cite{Hilborn} For a proper choice of initial conditions for $s(t)=0$ there is one excitation pulse before the system settles to the fixed point.

The concepts of membrane potential and ion currents lead naturally to thinking about these excitable tissues as electronic circuits. van der Pol and van der Mark suggested that the heart's dynamics can be modeled with a negative-resistance relaxation oscillator.\cite{van der Pol} Nagumo's circuit based on the Fitzhugh-Nagumo model has dynamics that are analogous to action potential propagation.\cite{Nagumo} References~\onlinecite{van der Pol} and \onlinecite{Nagumo} inspired many publications using electronic excitable circuits to investigate cardiac and neural systems. For the two circuits we present, a capacitor represents a piece of the cell membrane and the capacitor voltage is analogous to the membrane potential. Currents that charge the capacitor are analogous to inward membrane ion currents. Currents that discharge the capacitor are analogous to the outward potassium membrane currents. Voltage thresholds and time constants in the circuits account for the ion channels' voltage-gated behaviors and kinetics.

In the Fitzhugh-Nagumo circuit that we use (see Figs.~3 and 4) the capacitor voltage $V$ and inductor current $I$ are the variables analogous to $u$ and $v$ in Eq.~\eqref{fn_eqns}. The circuit block $\tilde V(V)$ produces voltage $\tilde V$ which is a cubic function of the voltage $V$. The input impedance to the $\tilde V(V)$ block is high so that the current at its input may be assumed to be zero. The differential equations for $V$ and $I$ are found by applying Kirchhoff's current law to the capacitor, and the voltage law to the branch containing the inductor. 
\begin{subequations}
\label{fn_circuit}
\begin{align}
\label{fn_circ_a}
C\frac{dV}{dt_s}&=\frac{\tilde{V}-V}{R} - I +I_s\\
\frac{dI}{dt_s}&=\frac{1}{L} (V-IR_L).
\end{align}
\end{subequations} 
Here $t_s$ indicates that the time variable has units. Equation~\eqref{fn_circuit} is made equivalent to Eq.~\eqref{fn_eqns} by using $s({t})=I_sR/2.5$, $b=R_L/R$, and converting to dimensionless variables $u$, $v$, $t$, and $\epsilon$ defined in Table~\ref{tab:conv}. The factor of 2.5 is the scaling factor of the MLT04 multiplier IC. The voltage $\tilde V$ is constructed so that $\tilde V-V$ corresponds to the nonlinear term $u(u-a)(1-u)$. Note that $\epsilon$ is the ratio of the time constants $RC$ and $L/R$ for charging the capacitor and establishing the inductor current, respectively. Because $\epsilon$ must be small for excitation pulses to occur, we obtain the conditions that charging the capacitor must be the fast process and creating the discharging current in the inductor is the slow one. The analogy to cardiac cells and neurons suggests that $u(u-a)(1-u)$ represents the inward ion currents and $v$ represents the outward ion currents. 

\begin{figure}[t]
 	\centering
		\includegraphics[width=0.35\textwidth]{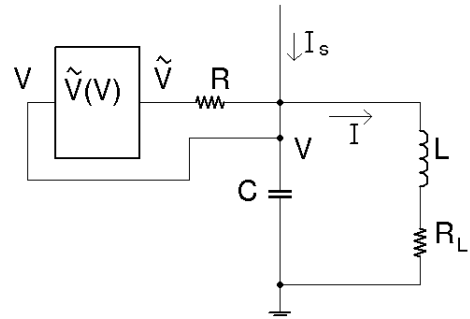} 
	\caption{\label{fig:fn_schem}Schematic of the Fitzhugh-Nagumo circuit. $R=100\,\Omega$, $C = 0.1\,\mu$f, and $L = 0.1$\,H. $R_L$ accounts for the inductor's intrinsic resistance (220\,$\Omega$), and an external resistor (33\,$\Omega$). The function block $\tilde V(V)$ is shown in Fig.~4.}
\end{figure}

\begin{figure*}
\centering
\includegraphics[width=0.8\textwidth]{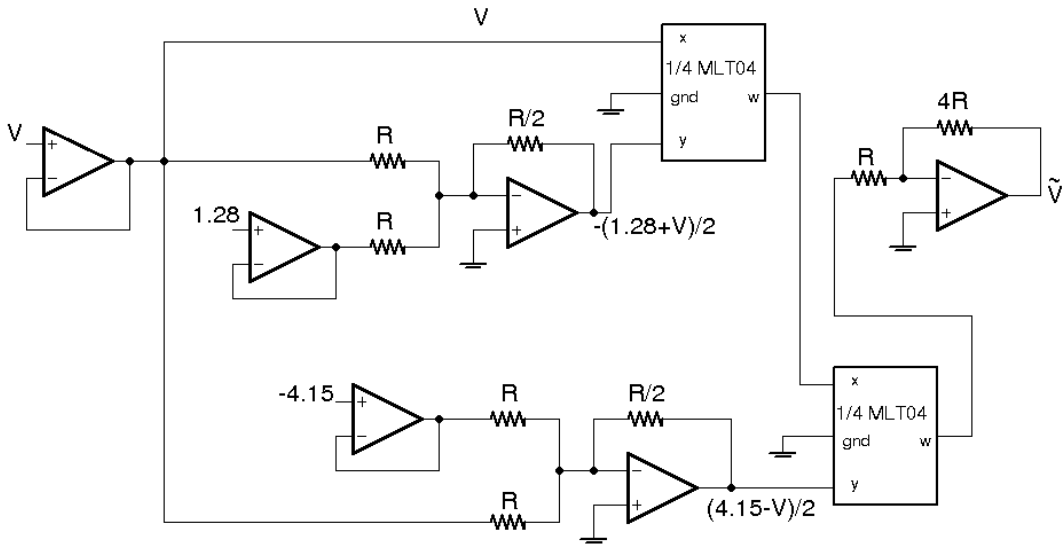}
\caption{\label{fig:function_circuit}Circuit for generation of function $\tilde V(V)$.  Only two of the MLT04's four multipliers are used. The unused multipliers (not shown) should be grounded.}
\end{figure*}

\begin{table}[h]
\begin{center}
\begin{tabular}{c|c|c}
Dimensionless& Fitzhugh-&\\
Quantity&Nagumo&\raisebox{1.5ex}{Three Transistor Circuit}\\
\hline
$u$ & $V/2.5$ & $V/5$\\
$v$ & $iR/2.5$ & $V_b/5$\\
$t$ & $t_s/RC$ & $t_s/R_fC$\\
$\epsilon$ & $R^2C/L$ & $R_fC/R_{\rm sl}C_{\rm sl}$\\
\hline
\end{tabular}
\caption{\label{tab:conv}Conversions from measured voltages and currents to dimensionless quantities. For the Fitzhugh-Nagumo circuit, Fig.~3 defines $V$, $I$, $R$, $C$, and $L$. For the three transistor circuit, Fig.~7(a) defines $V$, $V_b$, $R_f$, $C$, $R_{\rm sl}$, and $C_{\rm sl}$.}
\end{center}
\end{table}

Component values are chosen so that $\epsilon \approx 0.01$ and $b \approx 2.5$. To avoid the speed limitations of the op amps and the MLT04, we choose $RC=10\,\mu$s by using $R = 100\,\Omega$ and $C = 0.10\,\mu$f. The inductor has intrinsic resistance of 2.2\,$\Omega/$mH so an inductor resistance of $R_L=250\,\Omega$ corresponding to $b=2.5$ gives $L = 0.114$\,H. We include an external resistance in series with the inductor so that the inductor current can be measured easily from the voltage across this resistor. We use 33\,$\Omega$ which leaves about 220\,$\Omega$ for a nominal value $L=0.10$\,H; $R_L$ in Fig.~3 accounts for both the inductor's resistance and the external 33\,$\Omega$. 

The voltage $\tilde V$ is a function of $V$. Both voltages need to be normalized by the 2.5 scaling factor of the MLT04. With $u=V/2.5$ the required function for $\tilde u=\tilde V/2.5$ is
\begin{equation}
\label{uprime}
\tilde{u}=u(u-a)(1-u)+u =-u(u-r_1)(u-r_2).
\end{equation}
For $a=0.15$ the roots $r_1$ and $r_2$ are $1.66$ and $-0.512$. Figure 4 shows the circuit used to create the function $\tilde V(V)$ from Eq.~\eqref{uprime}. The roots are multiplied by 2.5 giving the values $-4.15$\,V and $1.28$\,V shown Fig.~4. Their associated polynomial factors are divided by 2 at the inverting addition amplifiers prior to the multiplier IC to bring the voltages within the $\pm 2.5$\,V range of the MLT04. Therefore the result after the multipliers is amplified by 4 at the final inverting amplifier. We use potentiometers to set the values $-4.15$ and $1.28$\,V, making it easy to change the roots in Eq.~\eqref{uprime}. Comte and Marqui\'{e} use a circuit with four fewer op amps to create a cubic polynomial.\cite{Comte} This reduction in circuit components comes at the expense of not being able to adjust the function so easily. They use their circuit to simulate nerve action potential propagation.\cite{Marq} 

We used inexpensive op amps (LF412) and multiplier IC (MLT04) to demonstrate that high precision components are not required to obtain the results shown in Fig.~1. Nominal component values were used for the results shown in Fig.~1. The measured capacitor voltage shown in Fig.~1 was normalized by the MLT04's scaling factor of 2.5. The inductor current was obtained from the voltage across the external $33\,\Omega$ and then converted to the dimensionless $v$ by multiplying by $100\Omega/(2.5$\,V). 

Nagumo's circuit based on the Fitzhugh-Nagumo equations uses a tunnel diode for the nonlinear element.\cite{Nagumo} We use the $\tilde V(V)$ block instead, because tunnel diodes are no longer inexpensive and readily available. The $\tilde V(V)$ block has the added benefit of allowing us to easily change the functional form of the nonlinear term. 

A constant current source for $I_s$ is provided by the circuit shown in Fig.~5. The current $I_s$ is the voltage drop across $R_E$ divided by $R_E$. The emitter voltage is approximately 0.65\,V above the base voltage due to the forward biased emitter-base junction. As an example, $R_1=20$\,k$\Omega$ and $R_2=10$\,k$\Omega$ gives a base voltage of 3.3\,V and emitter voltage of 4.0\,V. For $R_E = 1$\,k$\Omega$ the current is 1.0\,mA. These values correspond to $s(t)=I_sR/2.5=0.04$ in Eq.~\eqref{fn_eqns_a} for $R=100\,\Omega$ in Fig.~3. Use of a 25\,k$\Omega$ potentiometer for $R_2$ allows $s(t)$ to be in the range 0 to 0.085.

\begin{figure}[b]
\centering
\includegraphics[width=0.125\textwidth]{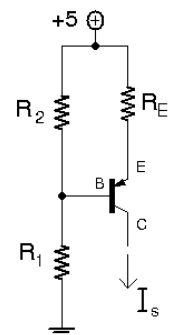}
\caption{\label{fig:isource}Constant current source $I_s$. $R_1=20$\,k$\Omega$, $R_2=10$\,k$\Omega$, and $R_E = 1$\,k$\Omega$ gives $I_s=1.0$\,mA. The transistor is a general purpose \textit{pnp} such as 2N3906.}
\end{figure}

A pulse train for the current source $I_s$ is provided by the circuit in Fig.~6 based on a 555 timer. The train's period and each pulse's width and amplitude are controlled by $R_a$, $R_b$, and $R_E$, respectively. For example, with $R_a=15$\,k$\Omega$, $R_b=470\,\Omega$, and $R_E=680\,\Omega$ the period, width, and height are approximately 1.1\,ms, 35\,$\mu$s, and 2.5\,mA, respectively. This current pulse corresponds to $s(t)$ making transitions between 0 and 0.1. Each pulse in this train initiates an excitation pulse in the Fitzhugh-Nagumo circuit. A reduction in the train period to 750\,$\mu$s results in ``missing" excitation pulses, a condition known as alternans. In cardiac research, alternans induced by external pacing provide information about heart function. 

\begin{figure}[t]
\centering
\includegraphics[width=0.35\textwidth]{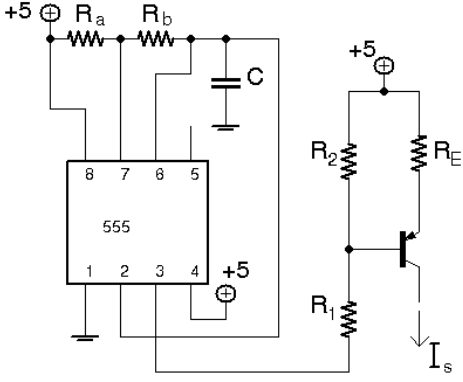}
\caption{\label{fig:pulser}Pulse train current source $I_s$. $R_a=100$\,k$\Omega$ potentiometer, $R_b=2$\,k$\Omega$ potentiometer, $C=0.1\,\mu$f, $R_1=R_2=22$\,k$\Omega$, and $R_E=1$\,k$\Omega$ potentiometer. The pulse period, width, and amplitude are controlled by $R_a$, $R_b$, and $R_E$, respectively.}
\end{figure}

Figure~3 looks similar to a RLC circuit. However its behavior is not that of a damped harmonic oscillator. In a harmonic oscillator, the current goes back and forth through the capacitor and inductor. In the Fitzhugh-Nagumo circuit the capacitor is charged by current through $R$ and discharged by current through $L$.

\section{Three Transistor Model}
A simple three-transistor circuit that has excitable behavior similar to the Fitzhugh-Nagumo system is shown in Fig.~7(a). The capacitor voltage $V$ and the transistor-base voltage $V_b$ are the excitatory and inhibitory variables analogous to $u$ and $v$. This circuit is much simpler than the Fitzhugh-Nagumo circuit in Figs.~3 and 4. It is possible to calculate all voltages and currents of the circuit (using SPICE for example). However, our goal is to use the excitable two-variable model in Eq.~\eqref{general}. Therefore, we model the circuit by the simpler circuit shown in Fig.~7(b) in which the two transistors responsible for the fast response positive feedback are replaced by the resistance $R_{\rm fast}(V)$ which is a function of the capacitor voltage $V$. The analysis of the circuit in Fig.~7(b) results in equations for the voltages $V$ and $V_b$,
\begin{subequations}
\label{3Tran}
\begin{align}
\label{3tranV}
C\frac{dV}{dt_s}&=\frac{5-V}{ R_{\rm fast}(V)}+\frac{5-V}{R_s}-\frac{V}{100k\Omega}\notag\\&-\frac{V-V_b}{ R_{\rm sl}}-I_C(V,V_b)\\
C_{\rm sl}\frac{dV_b}{dt_s}&=\frac{V-V_b}{R_{\rm sl}}-I_B(V,V_b).
\end{align}
\end{subequations}
Here, $I_C$ and $I_B$ are the collector and base currents for the transistor. This transistor provides the slow inhibitory response and is modeled using the Ebers-Moll equations.\cite{Ebers-Moll,Irwin} The slow response is determined by $R_{\rm sl}$ and $C_{\rm sl}$ at the transistor's base. The current $(5-V)/R_s$ is similar to the source $I_s$ in Eq.~\eqref{fn_circ_a}. If $R_s$ is removed ($R_s=\infty $), this is analogous to setting $I_s=0$, and $s(t)=0$ in Eq.~\eqref{fn_eqns_a}. A wide range of periods for self-exciting pulses can be obtained by varying $R_s$ from about 10 k$\Omega$ to 1 M$\Omega$. 

\begin{figure}[t]
	\begin{center}
		\subfigure[Complete circuit]{\label{fig:3Tran_schem-a}\includegraphics[width=0.3\textwidth]{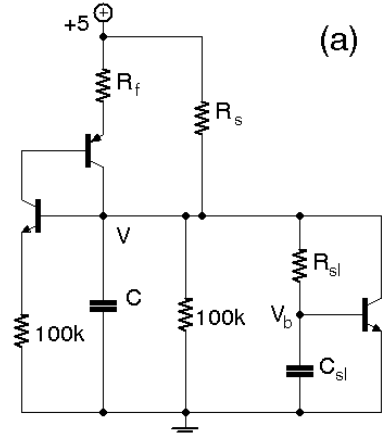}}
        \hspace{0.5in}
		\subfigure[Simple model]{\label{fig:3Tran_schem-b}\includegraphics[width=0.275\textwidth]{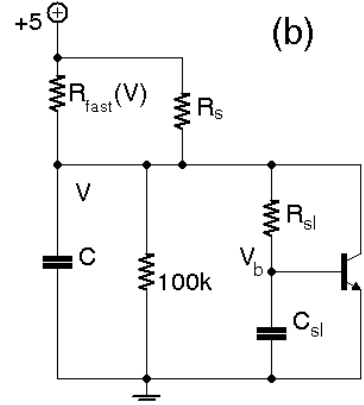}}
    \end{center}
    \caption{\label{fig:3Tran_schem}(a) Three-transistor excitable circuit and (b) its simplified model in which the two transistors responsible for the fast-response positive feedback are replaced by $R_{\rm fast}(V)$. $R_f=1$\,k$\Omega$, $C=0.33\,\mu$f, $C_{\rm sl}=1\,\mu$f, and $R_{\rm sl}=33$\,k$\Omega$. $R_s$ controls the period of self-exciting pulses. The \textit{npn} and \textit{pnp} transistors are 2N3904 and 2N3906 respectively.}
\end{figure}

Equation \eqref{3Tran} is put into the form of Eq.~\eqref{general} by using conversions in Table \ref{tab:conv}.  The result is
\begin{subequations}
\label{3Tran-dim}
\begin{align}
\label{3tranu}
\frac{du}{dt}&=(1-u)g_{\rm fast}(u)+
(1-u)\frac{R_f}{R_s}-u\frac{R_f}{100k\Omega}\notag\\&-(u-v)\frac{R_f}{R_{\rm sl}}-\frac{R_f}{5}I_C(u,v)\\
\frac{dv}{dt}&=\epsilon\Big[u-v-\frac{R_{\rm sl}}{5}I_B(u,v)\Big],
\end{align}
\end{subequations}
where $g_{\rm fast}(u)=R_f/R_{\rm fast}(V)$. Note that ${\epsilon}=0.01$ for our choice of circuit parameters, indicating that charging the capacitor is the fast process and increasing the transistor's base voltage (thereby increasing the discharging current) is the slow one. Note that the terms $g_{\rm fast}(u)$, $I_C(u,v)$, and $I_B(u,v)$ make Eq.~\eqref{3Tran-dim} more complicated than the Fitzhugh-Nagumo system, Eq.~\eqref{fn_eqns}.

\begin{figure}[t]
    \includegraphics[angle=0,width=0.45\textwidth]{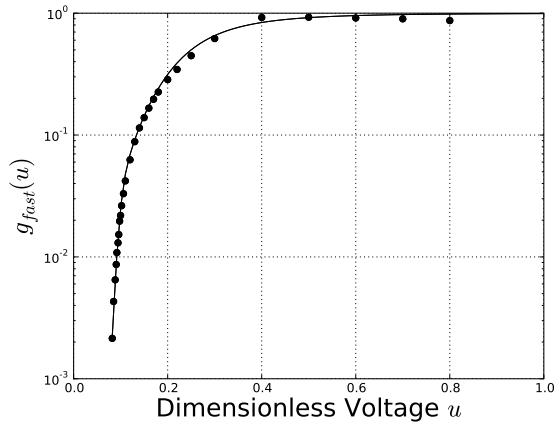}
    \caption{\label{fig:Gfast}Normalized measured conductance (filled circles) of the two-transistor fast-response positive feedback part of the three-transistor excitable circuit, and its calculated value (line) $g_{\rm fast}(u)$ from Eq.~\eqref{Gfast}. The conductance is normalized by 0.001\,${\Omega}^{-1}$, the conductance of $R_f$ in Fig.~7(a).}
\end{figure}
In the Appendix we show that the collector and base currents are
\begin{subequations}
\label{eq:ebers_ICall}
\begin{align}
\label{eq:ebers_IC}
I_C(u,v) &= -\frac{I_0}{ \beta_r}\big[e^{200(v-u)}-1\big]+I_0\big[e^{200v}-e^{200(v-u)}\big] \\
\label{eq:ebers_IB}
I_B(u,v)&=\frac{I_0}{\beta_f}\big[e^{200v}-1\big]+\frac{I_0}{\beta_r}\big[e^{200(v-u)}-1\big].
\end{align}
\end{subequations}
The SPICE values from the data sheet for a 2N3904 are $\beta_f=416$, $\beta_r=0.737$, and $I_0=6.7\times 10^{-15}$\,A. A suitable model for $g_{\rm fast}(u)$ is found by inspection of the fast-response portion of the circuit in Fig.~7(a). As $V$ increases from zero and passes through some threshold value, the two transistors turn on, allowing current to flow through the series combination of $R_f$ and the \textit{pnp} transistor. This current provides the threshold-triggered positive feedback that rapidly charges the capacitor. Thus $R_{\rm fast}(V)$ needs to have a steep transition from very large resistance down to $R_f$. The inverse of the resistance is the conductance, which must make a transition from zero to nearly $1/R_f$. We construct a function with these properties and fit it to data. The function we use is
\begin{equation}
\label{Gfast}
g_{\rm fast}(u)=\frac{R_f}{R_{\rm fast}(V)}=\frac{1}{\big[1+e^{w_1(V_{\rm th1}-5u)}\big]\big[1+(\frac{V_{\rm th2}}{5u})^{w_2}\big]}.
\end{equation}
Figure 8 shows Eq.~\eqref{Gfast} with the parameters $w_1=30$, $V_{\rm th1}=0.48$\,V, $w_2=3.5$, and $V_{\rm th2}=1.25$\,V, along with the experimentally determined conductance obtained by measuring the current through $R_f$ (and the \textit{pnp} transistor) as a function of the capacitor voltage $V=5u$. 

\begin{figure}[t]
	\centering
		\includegraphics[angle=0,width=0.45\textwidth]{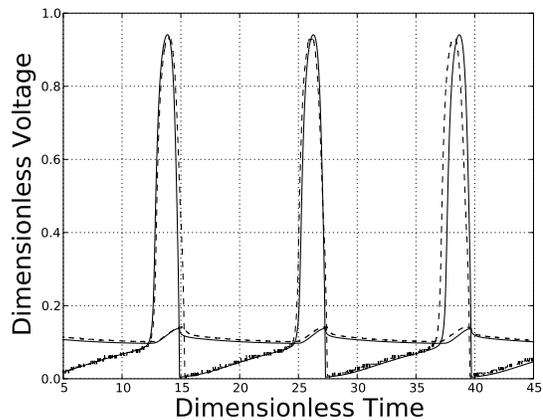}
	\caption{\label{fig:Trans_time}Numerical calculation (solid line) and circuit measurement (dashed line) for the three-transistor excitable circuit with $R_s = 330$\,k${\Omega}$. The excitatory variable $u$ demonstrates threshold-triggered positive feedback, rising slowly from $0$ to about $0.1$ where it then rises rapidly. The inhibitory variable $v$ undergoes smaller variations near $0.1$.}
\end{figure}

Figure 9 shows measurements of the normalized voltages $u$ and $v$ with $R_s=330$\,k$\Omega$, along with predictions using Eq.~\eqref{3Tran-dim}. The phase-space plot in Fig.~10 shows the reverse-$N$ shape for the $u$-nullcline that supports excitation pulses. The nullclines were found by setting the derivatives in Eq.~\eqref{3Tran-dim} to zero and numerically finding the zeros of $v$ as a function of $u$. 

\begin{figure}[h]
	\centering
		\includegraphics[angle=0,width=0.45\textwidth]{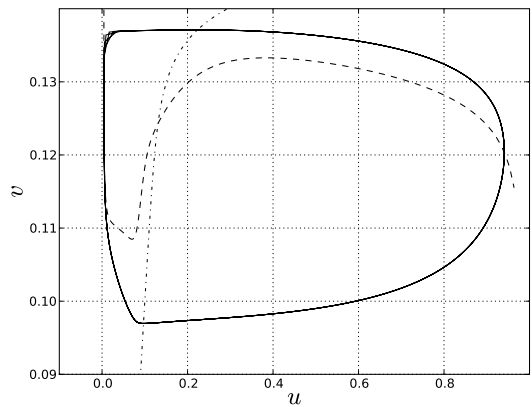}
	\caption{\label{fig:Trans_phase}Numerical calculation of the phase space plot (line) and nullclines ($u$-dash, $v$-dot-dash) for the three transistor circuit.}
\end{figure} 

\section{Application to Atrial Tachycardia}
As an example of the type of system that can be investigated with these circuits, we model atrial flutter, a tachycardia (fast heart rate) caused by reentrant circulation due to unidirectional block, and its treatment via surgical ablation. We include descriptions of the relevant physiology of the heart and the conditions leading to this arrhythmia. 

The self-exciting systems whose behavior is shown in Figs.~1 and 9 are analogous to the sinoatrial node, the natural pacemaker of the heart, located at the top of the right atrium. The source terms $s(t)$ and $(1-u)R_f/R_s$ in Eqs.~\eqref{fn_eqns_a} and \eqref{3tranu} correspond to the calcium ion currents that flow into cells of the sinoatrial node. These currents are responsible for the rising membrane potential which is the leading edge of the cardiac pacemaker pulse. The heartbeat signal generated by the sinoatrial node propagates through the contractile muscle cells of the atria, causing the atria to contract and send blood into the ventricles. Simultaneously, the signal travels through the modified, non-contractile cells of the cardiac conduction system to arrive at the atrioventricular node located between the right atrium and ventricle. Here the signal is delayed and ultimately sent down the specialized electrical conduction cells known as the Bundle of His to initiate the ventricular contraction. The delay allows time for blood to be emptied from the atria into the ventricles before contraction.

\begin{figure}[b]
\includegraphics[width=0.45\textwidth]{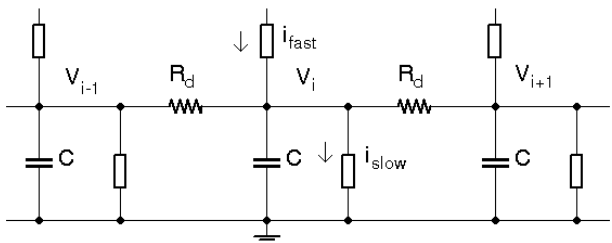}
\caption{\label{fig:propagate}Three coupled excitable cell circuits that simulate a 1-dimensional excitable medium. The characteristic time for diffusion between neighboring cells is $R_dC$.}
\end{figure}

Heartbeat signals (and nerve signals) propagate as action potentials in an excitable medium.  This medium is modeled by coupling excitable circuit cells together with resistor $R_d$ as shown in Fig.~11, similar to Nagumo's circuit based on tunnel diodes.\cite{Nagumo} Terms must be added to the right side of Eq.~\eqref{3tranV} [or Eq.~\eqref{fn_circ_a}] to account for current flow to and from neighboring circuit cells. These two terms are rearranged into the form of a diffusion term.
\begin{equation}
\frac{V_{i+1}-V_i}{R_d}-\frac{V_i-V_{i-1}}{R_d}=\frac{V_{i+1}-2V_i+V_{i-1}}{R_d} \to \frac{{\Delta x}^2}{R_d}\frac{\partial^2V}{\partial x^2},
\end{equation}
where $\Delta x$ is the distance between centers of neighboring cells of the medium. Diffusion is characterized by the diffusion coefficient $D$ where the characteristic time for diffusing the distance $\Delta x$ is ${\Delta x}^2/D$. Thus, the connection between diffusion and the electrical parameters is 
\begin{equation}
\frac{{\Delta x}^2}{D}=R_dC.
\end{equation}
The characteristic time for diffusion between neighboring cells corresponds to the $RC$ time constant for the capacitor and coupling resistor $R_d$. 

Because multiple excitable cells are required, we choose the simpler three transistor circuit in Fig.~7(a). Six of these cells are connected in a ring coupled by $R_d=47$\,k$\Omega$. A self-firing cell with $R_s=330$\,k$\Omega$ represents the sinoatrial node. The other cells do not have source resistor $R_s$. The cell on the ring opposite to the sinoatrial cell represents the atrioventricular node. The two sides of the ring represent two paths from the sinoatrial node through the atrial tissue to the atrioventricular node. Excitations started at the sinoatrial cell propagate down both sides of the ring and meet at the atrioventricular cell where they terminate because they cannot propagate through each other due to their refractory phases.

\begin{figure}[b]
\includegraphics[angle=0,width=0.45\textwidth]{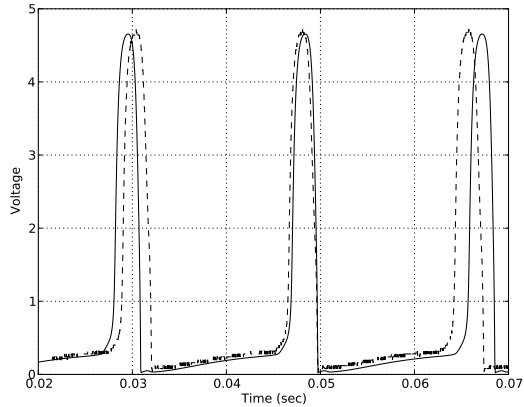}
\caption{\label{fig:normal}Numerical calculation (solid line) and measurement (dashed line) of the capacitor voltage at one cell on one side of the 6\,cell ring in which a source cell initiates pulses that propagate down both sides and meet at the cell opposite to source. The period is about 18\,ms. This behavior represents normal propagation of the heartbeat signal from the sinoatrial node through multiple paths merging at the atrioventricular node.} 
\end{figure}

The corresponding predictions provide a challenging computational physics project because they involve the numerical solution of partial differential equations. We use a simple Euler method with a time step $\Delta t = 2.5\,\mu$s chosen to be much smaller than the characteristic times for the positive feedback, $R_fC=(1$\,k$\Omega)(0.33\,\mu$f$)=330\,\mu$s and for diffusion, $R_dC=(47\,{\rm k}\Omega)(0.33\,\mu {\rm f})= 16$\,ms. Figure~12 shows the numerical predictions and data for voltage at a cell along one side of the ring as excitations generated at the sinoatrial cell propagate down both sides to the atrioventricular cell. The pulses terminate at the atrioventricular cell. Thus the measured period is the same as the firing period of the sinoatrial cell, about 18\,ms. 

\begin{figure}[t]
\includegraphics[width=0.35\textwidth]{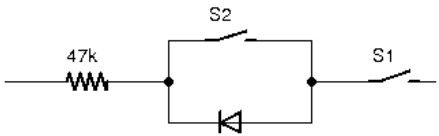}
\caption{\label{fig:uniblock}Coupling for either normal propagation (S1 and S2 closed), unidirectional block (S1 closed, S2 open), or ablation (S1 open). With S1 closed and S2 open pulse propagation from left to right is blocked, but from right to left still occurs.}
\end{figure}

Unidirectional block occurs when propagation is inhibited in the usual (orthograde) direction, but occurs for the backward (retrograde) direction. Thus, unidirectional block in one side of the ring results in a circulating pulse because the two pulses no longer meet and cancel at the atrioventricular cell. If the circulation period is less than the firing period of the sinoatrial cell, then the heartbeat rate at the atrioventricular cell will be higher than normal, which results in the tachycardia known as atrial flutter. In the ring circuit unidirectional block is implemented by including switches and a diode in series with one of the coupling resistors as shown in Fig.~13. Figure~14 shows our numerical predictions and data for the voltage at a cell for an excitation circulating around the 6\,cell ring due to unidirectional block. Note that the period is 11\,ms, less than for the normal signal propagation in Fig.~12. In reality, the atrial tissue may contain numerous paths of different lengths involved with reentrant dynamics. 

The normal beat period apparent in Fig.~12 is restored by breaking the circuit in the path containing the unidirectional block by opening switch S1 in Fig.~13. Breakage of the circuit corresponds to ablation (typically using radiofrequency heating) of the tissue containing the pathway with the unidirectional block. 
\begin{figure}[t]

\includegraphics[angle=0,width=0.45\textwidth]{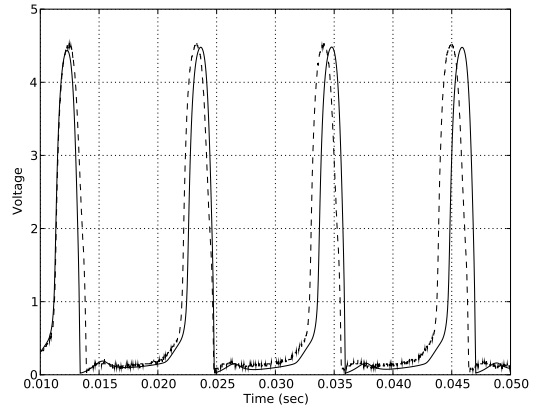}
\caption{\label{fig:circulate}Numerical calculation (solid line) and measurement (dashed line) of the capacitor voltage at one cell on a 6\,cell ring with circulating excitation caused by unidirectional block. The circulation period of about 11\,ms is shorter than in Fig.~12. The increased frequency due to circulating excitations is manifested as the reentrant tachycardia.}
\end{figure}

\section{Summary} 
We have presented two excitable systems, and an example of modeling reentrant tachycardia and its treatment by surgical ablation. These excitable systems may be used in a variety of explorations of excitable dynamics. Suggested possibilities include coupled self-firing excitable cells, external pacing (using Fig.~6) of an excitable media, and determining the minimum path length for an excitation circulating on a ring. 

\begin{acknowledgments}
This project was supported by an award from the Research Corporation. Matthew J.\ Lanctot did some preliminary work on this project. We thank Joseph Starobin for useful discussions.
\end{acknowledgments}

\appendix*

\section{The Ebers-Moll model for transistors}
The Ebers-Moll model for transistors is often unfamiliar to physics students. The advantage of this model is its applicability to all situations for the transistor, from fully off with base voltage equal to emitter voltage, all the way to fully on (saturation) with the collector voltage nearly equal to emitter voltage. Thus this transistor model covers both linear amplification and switching applications. We give a brief explanation of the model assuming familiarity with the basics of bipolar junction transistors. 

The standard current-voltage relation for a \textit{pn} semiconductor junction is 
\begin{equation}
\label{diode_eqn}
I(V)=I_0\Big[\exp\Big(\frac{V}{V_T} \Big)-1\Big],
\end{equation}
where $V_T= kT/e \approx 25$\,mV at room temperature and $I_0$ is a reverse saturation current. For an \textit{npn} bipolar junction transistor the base-emitter and base-collector form opposing \textit{pn} junctions. A voltage across one of the junctions causes current given by Eq.~\eqref{diode_eqn}. The transport factor $\alpha$ is the fraction of this current that diffuses across the base region thereby flowing through the second junction. In addition, the voltage across the second junction causes current at that junction given by Eq.~\eqref{diode_eqn}.  This current has an associated transport current at the first junction given by a transport factor for that direction of current flow. The result is that the transistor's collector and emitter currents each have two contributions, one similar to Eq.~\eqref{diode_eqn} due to the junction voltage at that terminal and the other due to the transport current associated with the voltage at the other junction. The difference between the collector and emitter currents is accounted for by the base current. Ebers and Moll show that the collector and emitter currents are\cite{Ebers-Moll}
\begin{subequations}
\label{eq:ebers}
\begin{align}
I_C=-\frac{I_0}{\alpha_r} \Big[\exp \Big(\frac{V_{BC}}{V_T} \Big)-1 \Big]+I_0 \Big[\exp \Big(\frac{V_{BE}}{V_T} \Big)-1\Big]\\
I_E=\frac{I_0}{\alpha_f} \Big[\exp \Big(\frac{V_{BE}}{V_T}\Big)-1\Big]-I_0 \Big[\exp\Big(\frac{V_{BC}}{V_T} \Big)-1 \Big].
\end{align}
\end{subequations} 
Each transport factor is related to the more familiar current gains $\beta_f$ and $\beta_r$ by $\beta = \alpha/(1-\alpha)$. A transistor is specified by $\beta_f$, $\beta_r$, and $I_0$. Equation~\eqref{eq:ebers} is used to express the collector and base currents for the slow-response transistor in Fig.~7(b) in terms of the capacitor voltage $V$, base voltage $V_b$, and the transistor parameters. The base current is $I_B=I_E-I_C$. Equation~ \eqref{eq:ebers_ICall} is obtained by noting that $V_{BE}=V_b$, $V_{BC}=V_b-V$, $1/V_T=40$\,V$^{-1}$, and converting to dimensionless $u$ and $v$.


\begin{thebibliography}{5}

\bibitem{Silverthorn}D. U. Silverthorn, W. C. Ober, C. W. Garrison, and A. C. Silverthorn, \textsl{Human Physiology, An Integrated Approach} (Prentice Hall, Upper Saddle River, NJ, 1998).

\bibitem{Fitzhugh}R. Fitzhugh, ``Impulses and physiological states in theoretical models of nerve membrane," Biophys. J. {\bf 1}, 445--466 (1961).

\bibitem{Nagumo}J. Nagumo, S. Arimoto, and S. Yoshizawa, ``An active pulse transmission line simulating nerve axon," Proc. Institute Radio Eng. {\bf 50}, 2061--2070 (1962). 

\bibitem{Bunton}P. H. Bunton, W. P. Henry, and J. P. Wikswo, ``A simple integrated circuit model of propagation along an excitable axon," Am. J. Phys. {\bf 64}, 602--606 (1996).

\bibitem{Yuan}G. Y. Yuan, S. G. Chen, and S. P. Yang, ``Eliminating spiral waves and spatiotemporal chaos using feedback signal," Eur. Phys. J. B {\bf 58}, 331--336 (2007).

\bibitem{Cassia}R. Cassia-Moura, F. G. Xie, and H. A. Cerdeira, ``Effect of heterogeneity on spiral wave dynamics in simulated cardiac tissue," Int. J. Bif. Chaos {\bf 14}, 3363--3375 (2004).

\bibitem{Starobin}I. B. Schwartz, I. Triandaf, J. M. Starobin, and Y. B. Chernyak, ``Origin of quasiperiodic dynamics in excitable media," Phy. Rev. E {\bf 61}, 7208--7211 (2000). 

\bibitem{Ross}M. H. Ross and W. Pawlina, \textsl{Histology A Text and Atlas} (Lippincott, Williams and Wilkins, Baltimore, MD, 2006).

\bibitem{Patel}A. Patel and S. M. Markowitz, ``Atrial tachycardia: mechanisms and management," Expert. Rev. Cardiovasc. Ther. {\bf 6}, 811--822 (2008).

\bibitem{HH}A. L. Hodgkin and A. F. Huxley, ``A quantitative description of membrane current and its application to conduction and excitation in nerve," J. Physiol. {\bf 117}, 500--544 (1952). 

\bibitem{van der Pol}B. van der Pol and J. van der Mark, ``The heartbeat considered as a relaxation oscillator, and an electrical model of the heart," Phil. Mag. {\bf 6}, 763--775 (1928).

\bibitem{Hilborn}R. C. Hilborn, \textsl{Chaos and Nonlinear Dynamics} (Oxford University Press, New York, 2003), 2nd ed., pp. 97--99. 

\bibitem{Comte}J. C. Comte and P. Marqui\'{e}, ``Generation of nonlinear current-voltage characteristics. A general method," Int. J. Bifurcat. Chaos {\bf 2}, 447--449 (2002).

\bibitem{Marq}P. Marqui\'{e}, J. C. Comte, and S. Morfu, ``Analog simulation of neural information propagation using an electrical Fitzhugh-Nagumo lattice," Chaos Soliton. Fract. {\bf 19} 27--30 (2004).

\bibitem{Ebers-Moll}J. J. Ebers and J. L. Moll, 
``Large signal behavior of junction transistors,'' Proc. Institute Radio Eng. {\bf 42}, 1761--1772
(1954).

\bibitem{Irwin}J. D. Irwin and D. V. Kerns Jr., \textsl{Introduction to Electrical Engineering} (Prentice Hall, Upper Saddle River, NJ, 1995), p. 333. 

\end{thebibliography}
\end{document}